\documentclass[twocolumn]{aa}

\usepackage{graphicx}
\usepackage{txfonts}
\usepackage{hyperref}
\usepackage{natbib}
\usepackage{soul}

\newcommand{\alfven}{Alfv\'en}
\newcommand{\vs}{{v_\mathrm{s}}}

\begin{document}

   \title{Multithermal apparent damping of slow waves due to strands with a Gaussian temperature distribution}
   \titlerunning{Multithermal apparent damping of slow waves}
%\shortauthors{Van Doorsselaere et al.}

   %\subtitle{I. Overviewing the $\kappa$-mechanism}

   \author{T. Van Doorsselaere
          \inst{1}
          \and
          S. Krishna Prasad\inst{2}
          \and
          V. Pant\inst{2}
          \and
          D. Banerjee \inst{2,3,4}
          \and
          A. Hood \inst{5}
          }

   \institute{Centre for mathematical Plasma Astrophysics, Department of Mathematics, KU~Leuven, Celestijnenlaan 200B, B-3001 Leuven, Belgium\\
              \email{tom.vandoorsselaere@kuleuven.be}
         \and
             Aryabhatta Research Institute of Observational Sciences, Nainital, India
             \and
             Indian Institute of Astrophysics, Koramangala, 560034, Bengaluru, India
             \and 
             Center of Excellence in Space Sciences, IISER, 741246, Kolkata, India
             \and 
             School of Mathematics and Statistics, University of St Andrews, St Andrews, Fife, KY16 9SS, UK
             }

   \date{Received ; accepted }

% \abstract{}{}{}{}{}
% 5 {} token are mandatory
 
  \abstract
  % context heading (optional)
  % {} leave it empty if necessary  
   {Slow waves in solar coronal loops are strongly damped. The current theory of damping by thermal conduction cannot explain some observational 
features.}
  % aims heading (mandatory)
   {We investigate the propagation of slow waves in a coronal loop built up from strands of different temperatures.   }
  % methods heading (mandatory)
   {We consider the loop to have a multithermal, Gaussian temperature distribution. The different propagation speeds in different strands lead to an 
multithermal apparent 
damping of the wave, similar to observational phase mixing. We use an analytical model to predict the damping length and propagation speed for the 
slow waves, including in imaging with filter telescopes.}
  % results heading (mandatory)
   {We compare the damping length due to this multithermal apparent damping with damping due to thermal conduction and find that the multithermal 
apparent damping is 
more important for shorter period slow waves. We have found the influence of instrument filters on the wave's propagation speed and 
damping. This allows us to compare our analytical theory to forward models of numerical simulations.}
  % conclusions heading (optional), leave it empty if necessary
   {We find that our analytical model matches the numerical simulations very well. Moreover, we offer an outlook for using the slow wave properties 
to infer the loop's thermal properties.}


   \maketitle
%
%________________________________________________________________

\section{Introduction} \label{sec:intro}
Since the turn of the century, slow waves in coronal loops are regularly observed through high resolution space observations \citep{berghmans1999}. 
These waves are seen as propagating intensity disturbances along open magnetic field or in the footpoint of loops 
\citep{demoortel2002,krishnaprasad2012}. In previous years, there was a debate on their interpretation in terms of slow waves or periodic flows 
\citep[e.g.][]{demoortel2015}, but for coronal loops or fans rooted in sunspots, there is a consensus that these are definitely slow waves driven by 
p-mode wave leakage in the sunspot \citep{banerjee2021}.\par

Slow waves in coronal loops are observed to be very heavily damped. Traditionally, it has been thought that the damping is caused by thermal 
conduction, after an extensive review of damping mechanisms by \citet{demoortel2003,demoortel2004}. Still, some of the observed properties of the 
damping may not be explained adequately by this traditional approach. For instance, in closed-field regions, it was found that the damping scales 
with 
the period with a positive coefficient \citep{krishnaprasad2014}, which is still more-or-less compatible with the damping by thermal conduction 
\citep{mandal2015}. However, for open-field regions in coronal holes, it was found that the damping scales with the period with a negative 
coefficient \citep{krishnaprasad2014,krishnaprasad2017}, which is incompatible with that damping theory. 
\citet{gupta2014} finds different damping behaviour in slow waves at different heights. At larger heights (10-70 Mm), they find shorter damping 
lengths for short period waves (as expected due to thermal conduction) but closer to the limb ($<$10 Mm), the long period waves ($>$6 min) appear to 
damp faster. Additionally, \citet{mandal2018} have shown, through a statistical study, that the damping length of slow waves in 
polar regions, indeed display a negative dependence on oscillation period. 
Moreover, \citet{krishnaprasad2019} find that the observed damping lengths of slow waves are much shorter than those expected from the theory of 
thermal conduction. Thus, it seems that other damping mechanisms are also at work. 
\par

Another point to consider is that the propagation speeds of slow modes in a loop seems to depend on the filter passband of the spacecraft that is 
used \citep{king2003,kiddie2012,uritsky2013}. This, too, seems to be incompatible with a slow mode wave propagating through a monolithic 
loop, subject to damping by thermal conduction. It was originally attributed to the fact that two adjacent loops (or aligned along the line-of-sight) 
would be observed in different filters. This may be correct, but it leaves the question why the slow waves are so coherently in phase between the 
observed structures.  \par

Another reason for disagreeing with the damping of slow waves by thermal conduction should be that it was previously observed that slow waves are 
damped with a Gaussian envelope \citep{krishnaprasad2014}. This feature could not be explained by damping by thermal conduction, nor resonant damping 
of slow waves in the cusp continuum \citep{yu2017a, yu2017b, geeraerts2022}. The latter is despite the fact that resonant absorption of kink modes 
in the \alfven{} continuum has been convincingly shown to result in a Gaussian damping profile \citep{pascoe2012, pascoe2017, pascoe2022}. However, 
the mechanism at work for resonant absorption in the \alfven{} continuum \citep{hood2013} does not seem to carry over to slow waves (Hood, 2015,
private communication). \par

Furthermore, it was argued by \citet{wang2015} that the thermal conduction coefficient is significantly suppressed in observed coronal loops. 
Despite the suppression of the thermal conduction, they still find a strong damping of slow waves. In their paper, they argue that this damping is 
caused by an enhanced compressive viscosity \citep{wang2019}. However, here we offer a suggestion that may also explain the observed strong damping 
of slow waves, without invoking unrealistically high viscosity coefficients \citep[a factor 10 higher than normally considered,][]{wang2019}. The 
model we propose below, does not need either thermal conduction or viscosity to result in apparent damping. \par

In DC heating models for the corona, it is thought that the loops are built out of isolated strands \citep[e.g.][]{aschwanden2000}, which are each
individually heated by nanoflares after which the higher temperature is redistributed by thermal conduction only along the magnetic field, but
inhibited across \citep[e.g.][and references therein]{williams2021}. They are consequently modelled as a collection of 1D field lines and their
thermodynamic evolution. This is in contrast to numerical evidence that loop strands have a short life time, because of the mixing by transverse
waves \citet{magyar2016}. The latter would result in a more continuous transverse temperature profile perpendicular to the magnetic field 
\citep{judge2023}. \par
In models of AC heating, transverse waves lead to turbulent behaviour in the loop boundary and entire cross-section \citep{karampelas2018},
leading to patchy heating in the cross-section, or in the turbulent layers \citep{vd2018,shi2021}. Despite the differences between AC and DC heating
models, it is safe to say that coronal loops do not have a uniform temperature across their cross-section. This will have a major impact on the
propagation properties of slow waves in those non-uniform temperature profiles. Here we will show that this leads to extra apparent damping (which we
call ``multithermal apparent damping'' or MAD) and different propagation speeds in different filter channels. In the future, this will allow us to 
infer the thermal
structure of coronal loops from the propagation and damping behaviour. 
Aside from this potential for probing the coronal thermal structure, nowadays slow waves are also considered for their ability to diagnose the 
coronal heating function \citep{kolotkov2019} through their perturbation of the energy balance equation for the background corona. Thus, it seems 
that slow waves are the optimal MHD waves for seismology of thermal
effects in the solar corona (perhaps aside from the entropy mode).

Even though our results were derived independently, it was pointed out in discussions at conferences that the physical effect we consider is the same
as was considered by \citet{voitenko2005}\footnote{In fact, given their pioneering idea on this, as much as 18 years before this manuscript, we may
consider naming the multithermal apparent damping as Voitenko-damping of slow waves.}. They modelled the propagation of sound waves in a 
multi-stranded loop
with strands drawn from a uniform distribution, as seen in a top-hat shaped instrumental filter. Here we go much beyond that initial description of
this phenomena.

\section{Results -- theoretical models}\label{sec:results}
We model a coronal loop as a superposition of strands, each with their own temperature and associated sound speed $\vs$. However, the model also 
carries over to a loop with a temperature continuously varying in its cross-section. Both of these models have no temperature variation along the 
magnetic field. In these models, a sound wave front is launched at the footpoint. We describe the propagation of the sound wave in the multithermal 
plasma.

\subsection{Intuition}\label{sec:intuition}
We take the $z$-direction along the uniform magnetic field, and only consider the hydrodynamic behaviour along the magnetic field lines
\citep[e.g.][]{demoortel2003,voitenko2005,mandal2015}. We
first think about the position $z_\mathrm{p}$ of the peak perturbation on the strands. We have that
\begin{equation} z_\mathrm{p}=z_0+\vs t, \end{equation}
where $z_0$ is the height of the initial excitation of the wave, in a strand with sound speed $\vs$. We consider the initial position of the peak 
$z_0$ to be independent of the strand, mimicking a joint impulsive excitation of the pulse low down in the atmosphere. 
In this paper, we consider a loop for which the strands have a sound speed that is randomly drawn from a normal distribution centred on $\bar{v}$ and 
standard deviation $\sigma_v$
\begin{equation}
	\vs\sim {\cal N}(\bar{v},\sigma_v^2).
\end{equation}
The distribution of the temperature in these strands is tightly related to the heating mechanism, which is currently not well understood 
\citep[e.g.][]{vd2020ssrv}. Thus, this assumption of a Gaussian distribution of strand's sound speed is an ad-hoc assumption in this paper. 
A sketch of the considered configuration is included in Fig.~\ref{fig:sketch}. \\ With such a Gaussian distribution of the strands, we then find that 
the peak positions $z_\mathrm{p}$ are also a Gaussian distribution. Following well-known rules for transforming random variables in 
statistics, we find that its distribution is
\begin{equation}
	z_\mathrm{p} \sim {\cal N}(z_0+\bar{v}t,t^2\sigma_v^2).\label{eq:gaussprop}
\end{equation}
We then consider all pulse perturbations on each strand to have the same amplitude. The line-of-sight integration over all these strands will then 
result in an intensity variation that is modelled well by this Eq.~\ref{eq:gaussprop}. This shows that the peak position distribution (and integrated 
intensity signal) propagates up with the average sound speed $\bar{v}$ in the loop bundle. It also shows that the peak 
position distribution steadily widens linearly in time, because its standard deviation goes as $t \sigma_v$. \par
The crucial realisation for understanding the multithermal apparent damping of sound waves is that the normalising factor of the Gaussian distribution 
with a
certain $\sigma$ is given as $1/\sigma \sqrt{2\pi}$. Applying this for the distribution of $z_\mathrm{p}$, we find that its peak value will vary over
time as
\begin{equation}
	\frac{1}{t\sigma_v\sqrt{2\pi}},
\end{equation}
and thus the wave will have a multithermal apparent damping that is proportional to $t^{-1}$. Indeed, given that there is no dissipation 
(all the wave energy $\int_\vs\int_z \rho v_z^2 d\vs dz$ is still in the infinitely long system), the damping is only apparent, because of the 
spreading of the wave front over time. This is because of the different propagation speeds in 
each strand, leading to an increasing spread in $z$. Thus, it is very similar to the process of phase mixing.

\subsection{Gaussian pulses}
\begin{figure}
    \centering
    \includegraphics[width=\columnwidth]{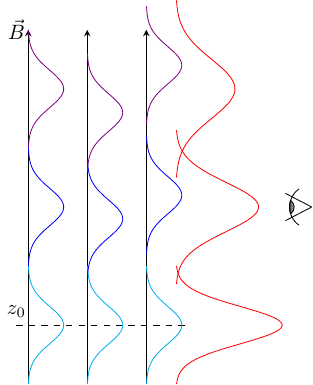}
    \caption{A schematic representation of the considered configuration. Three magnetic strands are shown. The cyan Gaussian pulses are excited at 
time $t=0$ at $z=z_0$ on all strands simultaneously. On each strand they propagate with a different speed, first to the blue line and then the purple 
line. The resultant observed intensity, as integrated over the different strands, is given by the red line, which shows the multithermal apparent 
damping and 
broadening.}
    \label{fig:sketch}
\end{figure}

Let us now build on this intuition to describe a system with an initial Gaussian pulse in the density perturbation, as could be the result of an 
impulsive excitation at the footpoint of the loop because of (e.g.) granular buffeting or a reconnection event. We imagine a group of strands all 
simultaneously excited with 
a pulse $W(z,0)$ of the form
\begin{equation}
	W(z,0)=a\exp{\left(-\frac{(z-z_0)^2}{2w^2}\right)}
\end{equation}
at position $z_0$, with pulse width $w$ and amplitude $a$. Because of the propagation of the pulse on each individual strand, at a later time, it 
will appear as
\begin{equation}
	W(z,t)=a\exp{\left(-\frac{(z-(z_0+\vs t))^2}{2w^2}\right)}, \label{eq:wavepacket}
\end{equation}
in which $\vs$ differs from strand to strand. A sketch of the configuration is shown in Fig.~\ref{fig:sketch}. \par

Now we look at the integrated signal for a Gaussian strand distribution for which $\vs\sim {\cal N}(\bar{v},\sigma_v^2)$ as before. We consider the 
integral of the different wavepackets (Eq.~\ref{eq:wavepacket}), with the distribution of the strand's sound speeds as weight function. So, in a 
sense, the integral is computing the intensity of a line-of-sight integration through the multistranded loop with a Gaussian distribution in the DEM 
\citep[differential emission measure, see e.g.][]{vd2018}. The emission measure is defined as $\int_z n^2 dz$ (with electron density 
$n$), which linearises to $2\int_z n_0 n_1 dz$ for background density $n_0$ and density perturbation $n_1$. The integral over the line-of-sight 
covers all strands with density $n_0$ in our loop model, and thus the wave perturbation $W(z,t)$ has to be multiplied with the Gaussian strand 
distribution. Moreover, the integral over $z$ (which traverses the entire loop system) is equivalent to integrating over all strands (in sound speed 
space).\par 

With this reasoning, the total signal $S(z,t)$ is then given as
\begin{equation}
	S(z,t)=\int_\vs \frac{1}{\sigma_v\sqrt{2\pi}} \exp{\left(-\frac{(\vs-\bar{v})^2}{2\sigma_v^2}\right)} a\exp{\left(-\frac{(z-(z_0+\vs 
t))^2}{2w^2}\right)} d\vs, \label{eq:superpos}
\end{equation}
or
\begin{equation}
	S(z,t)=\frac{a}{\sigma_v\sqrt{2\pi}} \int_\vs  \exp{\left(-\frac{1}{2}\left[\frac{(\vs-\bar{v})^2}{\sigma_v^2}+\frac{(z-(z_0+\vs 
t))^2}{w^2}\right]\right)} d\vs.
\end{equation}
Completing the square, the evaluation of the integral is given as
\begin{equation}
	S(z,t)=\frac{aw}{\sqrt{w^2+\sigma_v^2 t^2}} \exp{\left(-\frac{(z-(z_0+\bar{v} t))^2}{2(w^2+\sigma_v^2 t^2)}\right)}. \label{eq:final}
\end{equation}
This is a signal that peaks at $z_0+\bar{v} t$ and thus propagates with the average sound speed upwards. Its Gaussian width (as a function of $z$) is 
given by $\sqrt{w^2+\sigma_v^2 t^2}$, showing that it steadily increases in a hyperbolic fashion. For large $t$, the width increases approximately 
linearly.\par

As in Subsect.~\ref{sec:intuition}, we also note here that the amplitude of the peak signal (at $z=z_0+\bar{v} t$, i.e. co-propagating with the wave) 
decreases steadily over time. Its decay $d(t)$ from its initial amplitude is given as
\begin{equation}
	d(t)=\frac{w}{\sqrt{w^2+\sigma_v^2 t^2}}=\frac{1}{\sqrt{1+\frac{\sigma_v^2 t^2}{w^2}}}. \label{eq:damping}
\end{equation}
For large time $t$, this scales thus as $t^{-1}$, recovering the results of Subsect.~\ref{sec:intuition}. These latter results may also be recovered 
considering the limit of $w\to 0$, corresponding to an initial $\delta$-function perturbation.\par

\begin{figure}
	\centering
	\includegraphics[width=\columnwidth]{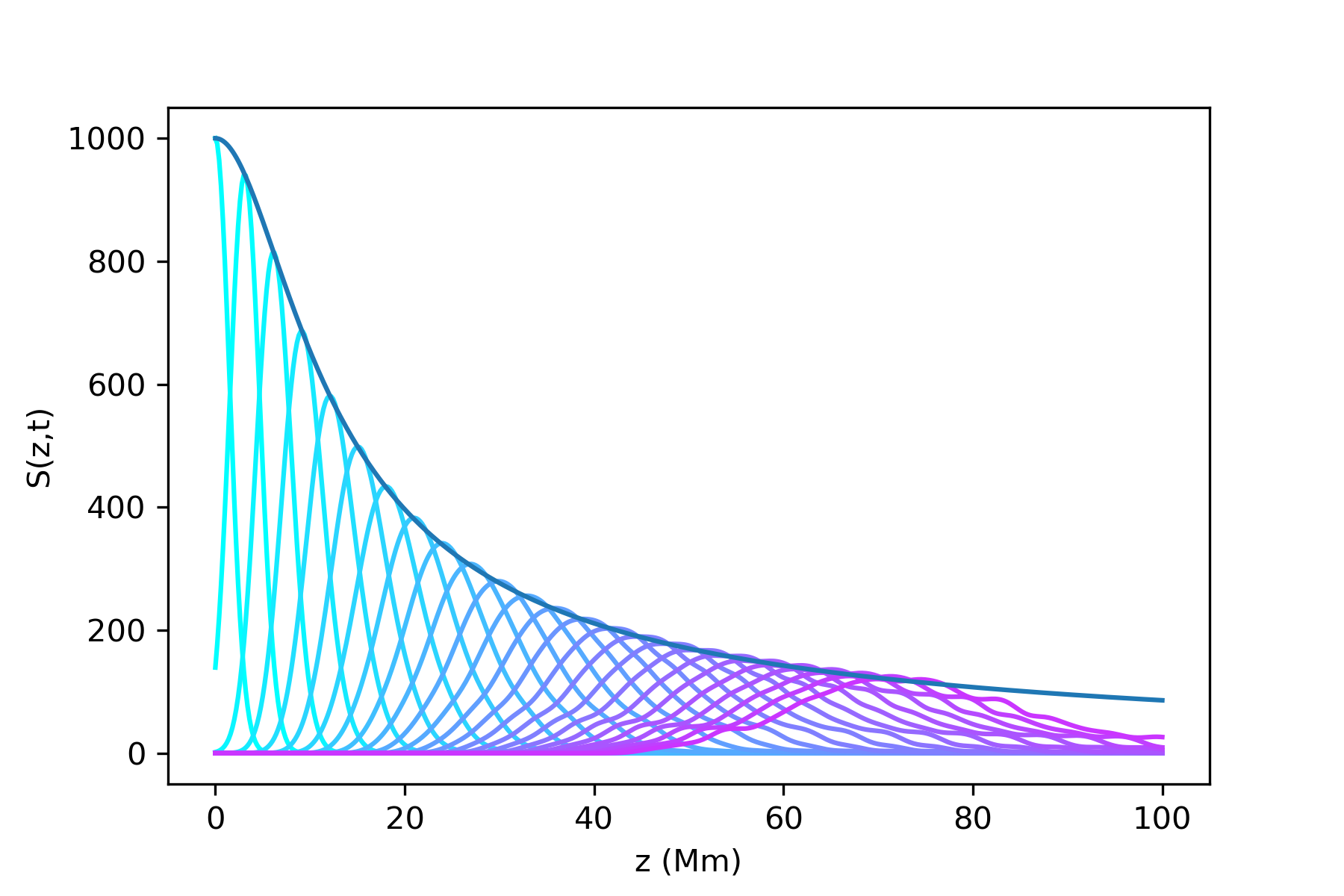}
	\caption{Comparison of Monte Carlo simulation with the analytical result. The analytically predicted envelope (Eq.~\ref{eq:damping}) is drawn with 
the dark blue line. The evolution of the Monte Carlo wave packet is drawn with the light blue to pink colour, progressively in time. The mean sound 
speed was taken as $\bar{v}=152$km/s and the spread in sound speed as $\sigma_v=26.4$km/s}
	\label{fig:compa_exp}
\end{figure}

In Fig.~\ref{fig:compa_exp}, we display the predicted damping envelope of Eq.~\ref{eq:damping}, and compare it to a Monte Carlo simulation of a 
Gaussian wave packet on different strands. For the Monte Carlo simulation, we have drawn 1000 $\vs$ from the normal distribution ${\cal 
N}(\bar{v},\sigma_v^2)$, with $\bar{v}=152$km/s (corresponding to 1MK) and $\sigma_v=26.4$km/s (for a motivation of these particular values, see 
Subsect.~\ref{sec:simulations}). The correspondence between the analytical solution and the Monte Carlo simulation is excellent. 

We may calculate the damping time $\tau$ as the e-folding time of this damping profile $d(t)$. We have then that
\begin{equation}
	e^{-1}=d(\tau)=\frac{1}{\sqrt{1+\frac{\sigma_v^2 \tau^2}{w^2}}}
\end{equation}
resulting in
\begin{equation}
	\frac{\sigma_v \tau}{w}=\sqrt{e^2-1}\approx 2.53 \quad \mbox{or}\quad \tau=\frac{w}{\sigma_v}\sqrt{e^2-1} . \label{eq:tau}
\end{equation}
\par

With a substitution $\Delta z\equiv z-z_0=\bar{v} t$, we may transform Eq.~\ref{eq:damping} to a damping profile as a function of $\Delta z$. We find 
that
\begin{equation}
	d(\Delta z)=\frac{1}{\sqrt{1+\frac{\sigma_v^2 \Delta z^2}{w^2 \bar{v}^2}}}.
\end{equation}
Making the same reasoning as in the derivation of the damping time $\tau$, we may also derive the damping length $L_\mathrm{d}$.
\begin{equation}
	e^{-1}=d(L_\mathrm{d})=\frac{1}{\sqrt{1+\frac{\sigma_v^2 L_\mathrm{d}^2}{w^2 \bar{v}^2}}}\quad \mbox{or} \quad L_\mathrm{d}=\frac{w 
\bar{v}}{\sigma_v}\sqrt{e^2-1}=\bar{v}\tau. \label{eq:length}
\end{equation}
\par

\subsection{Driven waves}\label{sec:driven}
\begin{figure}
	\centerline{\includegraphics[width=\columnwidth]{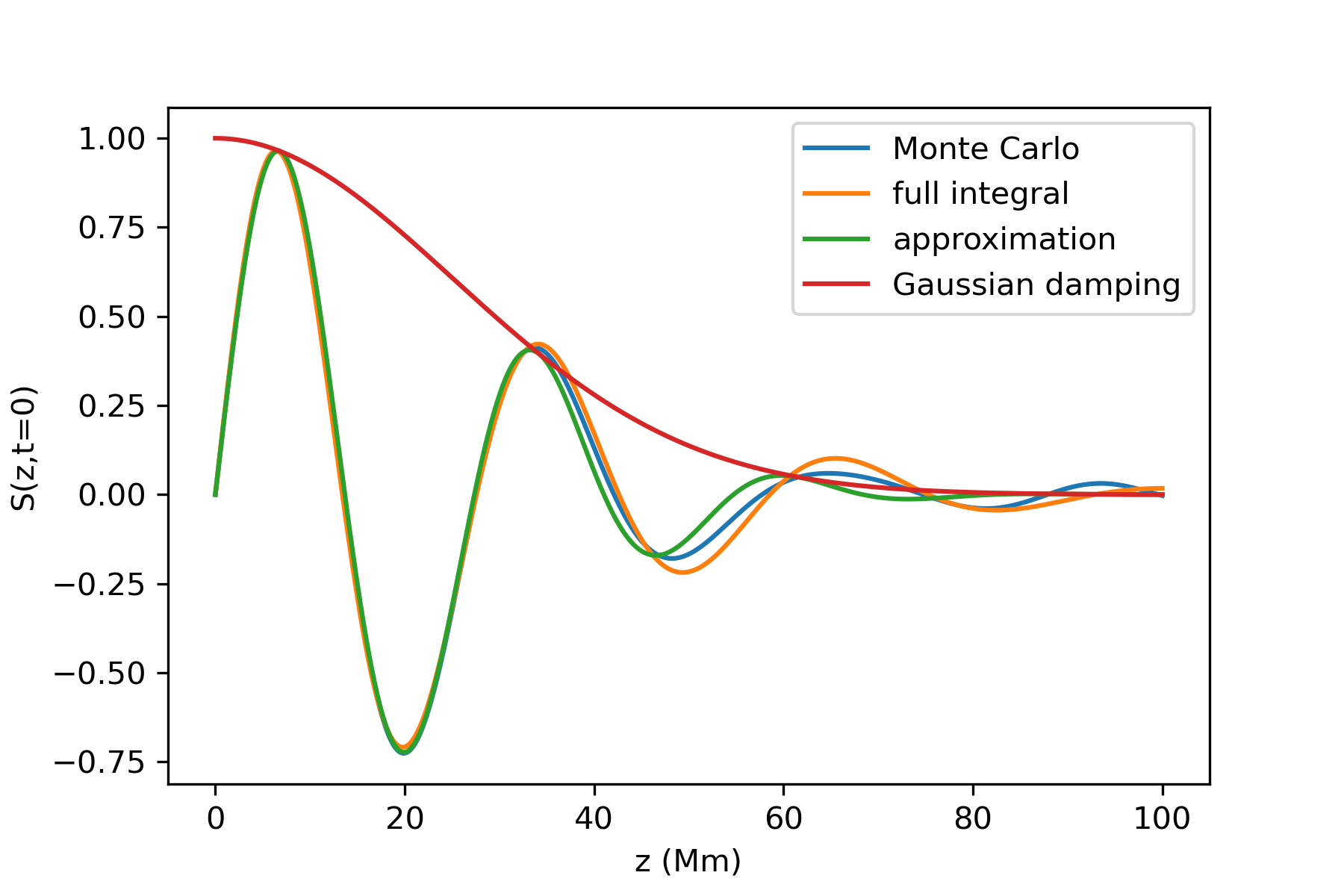}}
	\caption{Comparison of Monte Carlo simulation of driven sine functions (blue line) with the full integral (light orange) and the approximations 
in 
Eq.~\ref{eq:expcos} (green), all normalised to the starting value of 1. The expected Gaussian damping envelope (Eq.~\ref{eq:gaussdamp}) is shown in 
red. The time is arbitrarily chosen to be $t=0$, and the mean sound speed $152$km/s and spread in sound speed  $\sigma_v=26.4$km/s was chosen as 
before.}
	\label{fig:comparison_sine}
\end{figure}

Now we consider the case of driven, sinusoidal waves. At a certain height $z=0$, a periodic driver is inserted, resulting in propagating waves 
$a\sin{\left(kz-\omega t\right)}$, with amplitude $a$, frequency $\omega$ and wavenumber $k$. The resulting intensity signal of the loop 
bundle is then given as an integral (similar to Eq.~\ref{eq:superpos})
\begin{align}
	S(z,t)& =\int_\vs \frac{1}{\sigma_v\sqrt{2\pi}} \exp{\left(-\frac{(\vs-\bar{v})^2}{2\sigma_v^2}\right)} a\sin{\left(kz-\omega t\right)}
d\vs \nonumber \\ &=\frac{a}{\sigma_v\sqrt{2\pi}}\int_\vs  \exp{\left(-\frac{(\vs-\bar{v})^2}{2\sigma_v^2}\right)} \sin{\left(\frac{\omega
z}{\vs}-\omega t\right)}
d\vs, \label{eq:fullsine}
\end{align}
where we have used the dispersion relation $k=\omega/\vs$. The latter integral is not analytically solvable. We can still compare it to a Monte Carlo 
simulation with a 1000 $\vs$ drawn from a ${\cal N}(\bar{v},\sigma_v^2)$ distribution and summed up (see Fig.~\ref{fig:comparison_sine}). The Monte 
Carlo simulation is shown with the blue line, while the full integral in Eq.~\ref{eq:fullsine} is shown with the light orange line. The two lines 
closely 
match, and the deviation is due to the finite number of drawn $\vs$ values. For a higher number of draws, the two lines converge.

Further analytical progress is possible by making a Taylor approximation of the denominator in the sine:
\begin{align}
	\sin{\left(\frac{\omega z}{\vs}-\omega t\right)} & = \sin{\left(\frac{\omega z}{\bar{v}+\delta v}-\omega t\right)}\\
	&\approx \sin{\left(\frac{\omega z}{\bar{v}}(1-\frac{\delta v}{\bar{v}})-\omega t\right)}.
\end{align}
This approximation is valid if $\delta v \ll \bar{v}$. Since 95\% of the contribution to the full integral (Eq.~\ref{eq:fullsine}) is for $|\delta 
v|\leq 3\sigma_v$, the Taylor approximation is reasonably satisfied for our considered parameters of $\bar{v}=152$km/s and $\sigma_v=26.4$km/s, for 
which we subsequently have $|\delta v|\leq 3\sigma_v = 79.2\mbox{km/s} \lnapprox \bar{v}=152\mbox{km/s}$. So, the assumption $\delta v \ll \bar{v}$ 
seems to be sufficiently well satisfied in loops that are not too extremely multithermal (i.e. with $\sigma_v\lesssim \bar{v}$). This Taylor 
approximation allows to rewrite 
$S(z,t)$ as
\begin{multline}
	S(z,t)\approx \frac{a}{\sigma_v\sqrt{2\pi}} \left\lbrace \sin{\left(\frac{\omega z}{\bar{v}}-\omega t\right)} \int_{\delta v}  
\exp{\left(-\frac{\delta v^2}{2\sigma_v^2}\right)} \cos{\left(\frac{\omega z \delta v}{\bar{v}^2}\right)} d\delta v \right. \\ \left. -
\cos{\left(\frac{\omega
z}{\bar{v}}-\omega t\right)} \int_{\delta v}  \exp{\left(-\frac{\delta v^2}{2\sigma_v^2}\right)} \sin{\left(\frac{\omega z \delta 
v}{\bar{v}^2}\right)} d\delta v \right\rbrace.
\end{multline}
It turns out that the rightmost integral in this expression is exactly 0, because its integrand is an odd function in $\delta v$. Thus, we have that
\begin{equation}
	S(z,t)\approx \frac{a}{\sigma_v\sqrt{2\pi}} \sin{\left(\frac{\omega z}{\bar{v}}-\omega t\right)} \int_{\delta v}  \exp{\left(-\frac{\delta 
v^2}{2\sigma_v^2}\right)} \cos{\left(\frac{\omega z \delta v}{\bar{v}^2}\right)} d\delta v . \label{eq:expcos}
\end{equation}
The numerically calculated result of Eq.~\ref{eq:expcos} is shown in Fig.~\ref{fig:comparison_sine} with the green line. It matches the Monte Carlo 
simulations (blue) and full integral (light orange) reasonably well. The integral in Eq.~\ref{eq:expcos} can be calculated analytically by 
writing it as a complex function: %\footnote{Following the answer at Stackexchange: \url{https://math.stackexchange.com/a/578452}}:
\begin{equation}
	\int_{\delta v}  \exp{\left(-\frac{\delta v^2}{2\sigma_v^2}\right)} \cos{\left(\frac{\omega z \delta v}{\bar{v}^2}\right)} d\delta v = \Re 
\int_{\delta v} \exp{\left(-\frac{\delta v^2}{2\sigma_v^2}+\imath \frac{\omega z \delta v}{\bar{v}^2}\right)}d\delta v.
\end{equation}
We subsequently have
\begin{align}
	& \Re \int_{\delta v} \exp{\left(-\frac{\delta v^2}{2\sigma_v^2}+\imath \frac{\omega z \delta v}{\bar{v}^2}\right)}d\delta v \\ &= \Re
\int_{\delta v}
\exp{\left(-\left[\frac{\delta v}{\sqrt{2}\sigma_v}-\imath \frac{\sqrt{2}\sigma_v \omega 
z}{2\bar{v}^2}\right]^2-\frac{\sigma_v^2\omega^2}{2\bar{v}^4}z^2\right)} d\delta v\\
	&= \exp{\left(-\frac{\sigma_v^2\omega^2}{2\bar{v}^4}z^2\right)} \Re \int_{\delta v} \exp{\left(-\left[\frac{\delta v}{\sqrt{2}\sigma_v}-\imath 
\frac{\sqrt{2}\sigma_v \omega z}{2\bar{v}^2}\right]^2 \right)} d\delta v \\
	&= \sqrt{2\pi} \sigma_v \exp{\left(-\frac{\sigma_v^2\omega^2}{2\bar{v}^4}z^2\right)}.
\end{align}
Inserting this in Eq.~\ref{eq:expcos}, we find as end result
\begin{equation}
	S(z,t)\approx a\exp{\left(-\frac{\sigma_v^2\omega^2}{2\bar{v}^4}z^2\right)} \sin{\left(\frac{\omega z}{\bar{v}}-\omega t\right)} . 
\label{eq:signalsine}
\end{equation}
The wave is thus propagating with the average sound speed, and has additionally a Gaussian damping envelope with a Gaussian damping length
\citep[keeping the traditional factor 2 in the denominator,][]{pascoe2017}
\begin{equation}
	L_G=\frac{\bar{v}^2}{\sigma_v\omega}. \label{eq:gaussdamp}
\end{equation}
For the values considered in this paper ($\bar{v}=152$km/s, $\sigma_v=26.4$km/s, $\omega=2\pi/180$s), this reduces to a damping length $L_G=25.1$Mm.

\begin{figure}
	\centerline{\includegraphics[width=\columnwidth]{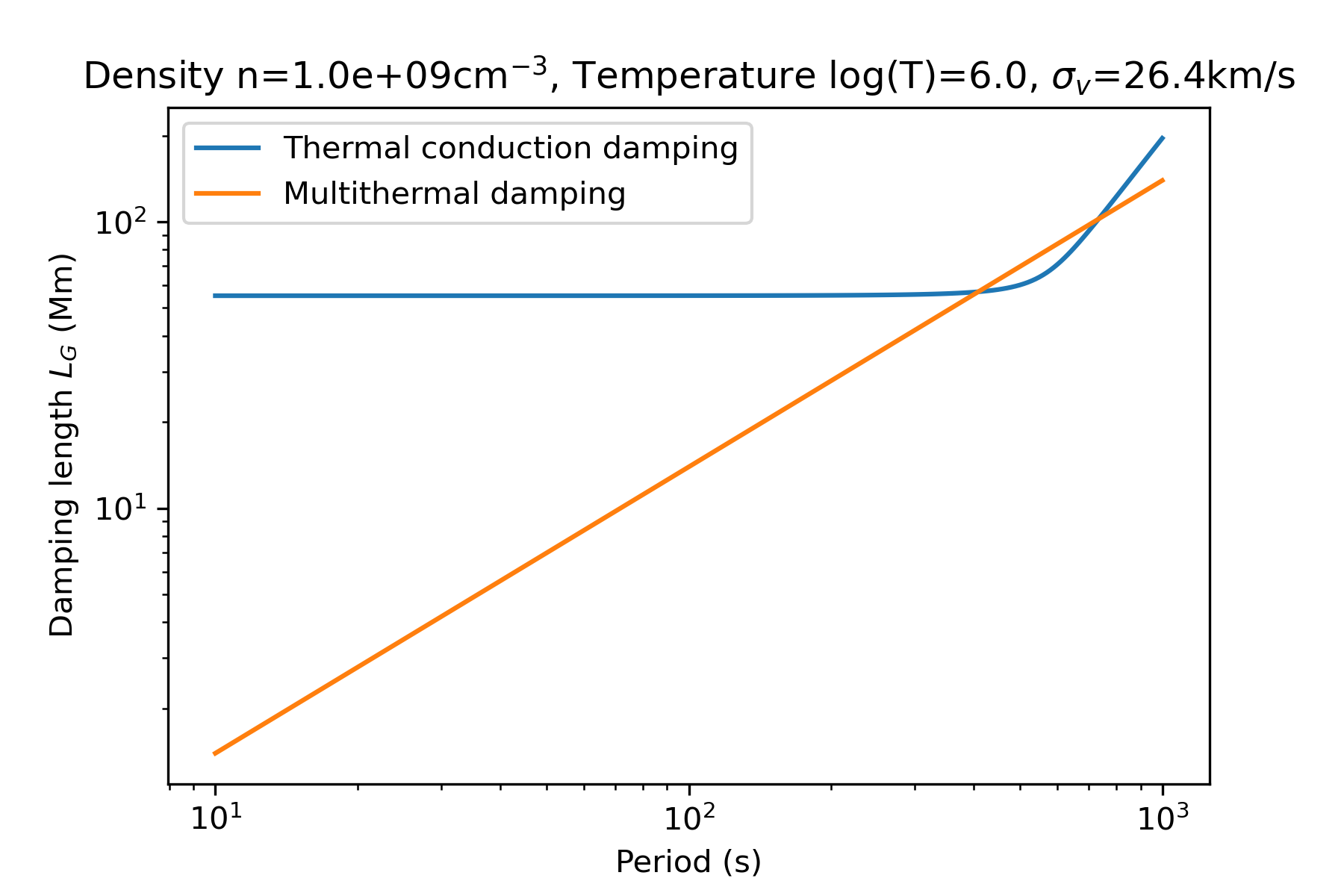}}
	\caption{Here we show the expected damping lengths (in Mm) as a function of period (in s), for both multithermal apparent damping 
(Eq.~\ref{eq:gaussdamp}) 
and thermal conduction \citep[Eq.~3 in][]{mandal2015}. The density was taken to be $10^9\mathrm{cm}^{-3}$, the mean temperature as $10^6$K, and the 
spread in temperature as $\sigma_v=26.4$km/s.}
	\label{fig:comparison}
\end{figure}

The formula shows that the damping length is inversely proportional to the frequency. This is a different dependence than the thermal conduction 
damping length, which is proportional to $\omega^{-2}$. In Fig.~\ref{fig:comparison}, we compare the multithermal apparent damping to the damping by 
thermal 
conduction. For the latter, we have taken the results in \citet{mandal2015}, and these are shown with the blue line. The other, light orange line 
corresponds to Eq.~\ref{eq:gaussdamp}. The graph shows that for intermediate periods (i.e. between 300s and 1000s), the damping by thermal conduction 
is comparable.
However, for shorter or longer periods, the multithermal apparent damping becomes more significant. Caution is appropriate here, because the 
multithermal  apparent
damping has a
Gaussian damping profile which is compared in this graph to the exponential damping profile of the thermal conduction.

\section{Results -- loops in filter images}

\subsection{Influence of a finite filter in imaging observations}\label{sec:filter}
Let us consider the influence of a filter on the observability and multithermal apparent damping of slow waves. For a filter $F$ described by a 
Gaussian 
function in $\vs$-space as
\begin{equation}
	F(\vs)=a_\mathrm{F}\exp{\left(-\frac{(\vs - v_\mathrm{F})^2}{2\sigma_\mathrm{F}^2}\right)}
\end{equation}
with amplitude $a_\mathrm{F}$, mean $v_\mathrm{F}$ and width $\sigma_\mathrm{F}$, the resulting observed signal (equivalent to 
Eq.~\ref{eq:superpos}) would be
\begin{multline}
	S(z,t)=\int_\vs \frac{1}{\sigma_v\sqrt{2\pi}} \exp{\left(-\frac{(\vs-\bar{v})^2}{2\sigma_v^2}\right)} a_\mathrm{F}\exp{\left(-\frac{(\vs - 
v_\mathrm{F})^2}{2\sigma_\mathrm{F}^2}\right)} \\ a\exp{\left(-\frac{(z-(z_0+\vs t))^2}{2w^2}\right)} d\vs.
\end{multline}
The first two Gaussian distributions may be combined by realising that
\begin{multline}
	\exp{\left(-\frac{(\vs-\bar{v})^2}{2\sigma_v^2}\right)} a_\mathrm{F}\exp{\left(-\frac{(\vs - v_\mathrm{F})^2}{2\sigma_\mathrm{F}^2}\right)} \\ =
a_\mathrm{F} \exp{\left(-\frac{(v_\mathrm{F}-\bar{v})^2}{2(\sigma_\mathrm{F}^2+\sigma_v^2)}\right)} \exp{\left(-\frac{(\vs-V)^2}{2\Sigma^2}\right)}, 
\label{eq:filter}
\end{multline}
where we have introduced the notation
\begin{equation}
	\frac{1}{\Sigma^2}=\frac{1}{\sigma_\mathrm{F}^2}+\frac{1}{\sigma_v^2} \qquad V=\frac{\sigma_v^2v_\mathrm{F} + \sigma_\mathrm{F}^2 
\bar{v}}{\sigma_\mathrm{F}^2+\sigma_v^2} \label{eq:filterspeed}
\end{equation}
for the width $\Sigma$ and average $V$ of the resulting Gaussian. This means that the width of the Gaussian is always decreased, due to the harmonic 
average. \par
These expressions for $\Sigma$ and $V$ may then be inserted in Eq.~\ref{eq:final}, while also remembering to also take the extra factors of 
Eq.~\ref{eq:filter} along and incorporate them with $a$. This will result in
\begin{equation}
	S(z,t)=\frac{aa_\mathrm{F} w}{\sqrt{w^2+\Sigma^2 t^2}} \exp{\left(-\frac{(v_\mathrm{F}-\bar{v})^2}{2(\sigma_\mathrm{F}^2+\sigma_v^2)}\right)} 
\exp{\left(-\frac{(z-(z_0+V t))^2}{2(w^2+\Sigma^2 t^2)}\right)}.
\end{equation}
As before, the damping (following a wave packet, at a ray of $z=z_0+Vt$) will be as
\begin{equation}
	d(t)=\frac{w}{\sqrt{w^2+\Sigma^2 t^2}}.
\end{equation}
This is weaker damping than in the non-filtered case, because $\Sigma<\sigma_v$.\\
Likewise, we may also insert $\Sigma$ and $V$ for $\sigma_v$ and $\bar{v}$ respectively in the Gaussian damping lengths (Eq.~\ref{eq:gaussdamp}):
\begin{equation}
	L_G=\frac{V^2}{\Sigma\omega}. \label{eq:gaussobs}
\end{equation}

It is also possible to use the propagation speed in different filters to estimate the temperature spread $\sigma_v$ and the mean temperature 
$\bar{v}$. This model naturally explains the different propagation speeds in different filter channels and this difference may be used to measure the 
loop's fundamental thermal properties and quantify its DEM. As in Eq.~\ref{eq:filterspeed}, we see that $\sigma_\mathrm{F}$ and $v_\mathrm{F}$ are 
known for each filter. Then the propagation speed $V$ may be measured in different filters, allowing us to estimate $\bar{v}$ and $\sigma_v$ through 
least squares fitting.\\
Moreover, the effective filter width $\sigma_\mathrm{F}$ is much larger for imaging observations than spectral observations. Thus, it is to be 
expected that imaging observations are much more profoundly impacted by the effect of multithermal apparent damping. Spectral observations will 
experience 
very little damping from the multithermal apparent damping effect, given the narrowness of the effective filter. Thus, we may use the combination and 
contrast 
between imaging and spectral observations to disentangle the real damping mechanism from the multithermal apparent damping. That opens exciting 
prospects for 
seismology of thermal properties of coronal loops. 

\subsection{Comparison with simulations}\label{sec:simulations}
%{\color{red} KP: text on Krishna's simulations}
Here, we verify the analytical results described in the previous sections using a 3D MHD multithermal  loop model similar to the one presented in
\citet{krishnaprasad2023}. They solve the ideal MHD equations with MPI-AMRVAC \citep{porth2014}, where only numerical diffusion is present and no 
explicit diffusive terms.  They consider a bundle of 33 vertical strands with randomly
assigned plasma temperatures and densities to represent a coronal loop, similar to the setup we consider in this paper. To elaborate, the plasma 
temperature $T$ (number density $n$) for each strand
is selected from a random normal distribution whose peak value corresponds to log{\,}$T=6.0$ (log{\,}$n=9.2$), with a standard width of 0.15 (0.10).
The plasma temperature outside the strands and that outside the loop are kept at the same value, 1{\,}MK. The corresponding number densities are fixed
at 5$\times$10$^8${\,}cm$^{-3}$. The magnetic field is vertical and parallel to the axis of the loop. For further details on the simulation setup, we
refer the interested reader to \citet{krishnaprasad2023}. \par

\begin{figure}
	\centerline{\includegraphics[width=\columnwidth]{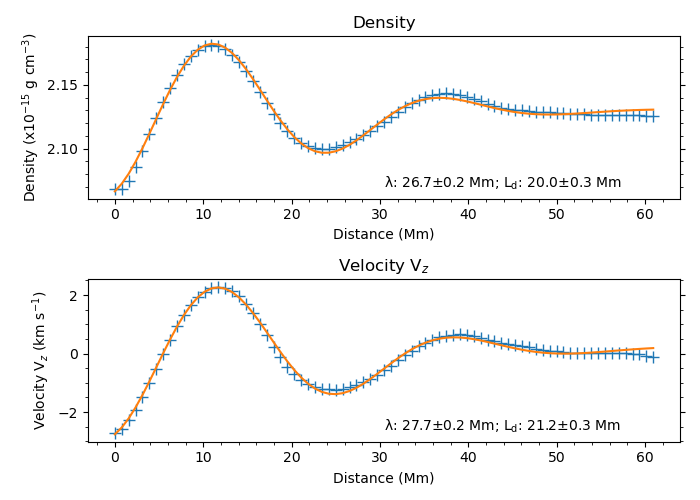}}
	\caption{The average density (top) and vertical velocity, $v_z$ (bottom) profiles along the simulated multithermal loop. The solid lines represent
a Gaussian damped sinusoid fit to the data following the function given in Eq.{\,}\ref{eq:gdamp}. The obtained wavelength ($\lambda$) and damping
length $L_d$ values are listed in the plot.}
	\label{fig:dlen}
\end{figure}

Considering the peak ($\mu_{\log{T}}=6.0$) and the width ($\sigma_{\log{T}}=0.15$) values of temperature distribution in the simulations, and assuming
that the resulting distribution of sound speeds is sufficiently normal (so that the theory in Sec.~\ref{sec:results} applies), we can estimate the
sound speed distribution properties from the temperature distribution. For that, we calculate that $\log{T}=6.0\pm 0.15$ corresponds to a sound speed
value of $\vs=152^{+28}_{-24}\mathrm{km/s}$, by calculating the sound speed for $\mu_{\log{T}}$ and $\mu_{\log{T}}\pm\sigma_{\log{T}}$ separately. So,
in what follows we take the following parameters: $\bar{v}=152\mathrm{km/s}$ and $\sigma_v=26.4\mathrm{km/s}$. Since the period of the driver in the
simulation was 180s, this $\bar{v}$ is expected to result in a wavelength of 27.4Mm. The wavelength values $\lambda$ obtained in
Fig.{\,}\ref{fig:dlen} indicate a good agreement with this.  \par

In their model, \citeauthor{krishnaprasad2023} excite slow magneto-acoustic waves within the loop by periodically (period $\approx$180{\,}s)
driving the vertical velocity ($v_z$) at the 
bottom boundary of the loop, with an amplitude of $\approx$7.6{\,}km{\,}s$^{-1}$ which is approximately 5\% of the sound speed at 1MK. This 
driving amplitude was chosen to be small, because we wanted to avoid any damping caused by non-linear effects of the waves.  We use the same driver
in this study, however, to highlight the multithermal effects we restrict the spatial location of the driver to the positions of the strands. In other
words, the amplitude of the driver is zero outside the strand locations and consequently the oscillations are restricted to the strands. Once the
generated slow waves start approaching the top boundary of the loop, we compute the mean density and vertical velocity ($v_z$) across the loop as a
function of distance along the loop to analyse how the oscillation amplitudes evolve.
Fig.{\,}\ref{fig:dlen} displays the density and $v_z$ profiles along the loop in the top and bottom panels respectively. As can be seen, the 
oscillations appear to damp very quickly. For a proper quantitative assessment, we measure the damping lengths by fitting the data with the following 
damped sinusoid function 
\begin{equation}
\label{eq:gdamp}
f(z) = A_0{\,}\mathrm{exp}\left(\frac{-z^2}{2 L_d^2} \right){\,}\mathrm{sin} \left(\frac{2\pi z}{\lambda}+\phi \right) +b_1 z +b_0.
\end{equation}
Here $z$ is the coordinate along the loop axis, $A_0$ is the maximum amplitude, $L_d$ is the damping length, $\lambda$ is the wavelength, $\phi$
is the initial phase, and $b_1$ and $b_0$ are appropriate constants. It may be noted that this function describes a Gaussian damped sine wave similar
to that described in \citet{pascoe2016}. Although an exponential damping is generally considered for slow waves, as described in
Sec.~\ref{sec:driven}, the multithermal apparent damping is expected to produce a Gaussian damping which justifies our choice here. The solid light 
orange
lines in Fig.{\,}\ref{fig:dlen} represent the obtained fits to the data. The damping lengths obtained from the fits are 20$\pm$0.3{\,}Mm, and
21.2$\pm$0.3{\,}Mm for the density and vertical velocity respectively. These values are within 20\% of the expected value of 25Mm (see
Sec.~\ref{sec:driven}, Eq.~\ref{eq:gaussdamp}), and are thus a reasonable match. The deviation may be due to (1) the approximation of the full
integral (Eq.~\ref{eq:fullsine} by Eq.~\ref{eq:expcos}), or (2) the ``small'' number of strands (only 33) in the simulations of
\citet{krishnaprasad2023}, insufficient to fully cover the continuous, Gaussian DEM which is modelled in Eq.~\ref{eq:fullsine} due to the finite
sample size. Because of the chosen small driver amplitude, non-linear effects do not play a role in the damping of these waves. Numerical diffusion 
could play a role, but we have verified that the increasing the numerical resolution has no effect on the measured damping lengths. \par

For a direct comparison with observations, we also forward model the data using the FoMo code \citep{vd2016fomo}. In particular, we generate synthetic
images in the 6 coronal channels of SDO/AIA, namely, 94{\,}{\AA}, 131{\,}{\AA}, 171{\,}{\AA}, 193{\,}{\AA}, 211{\,}{\AA}, and 335{\,}{\AA}. As
described in \citet{krishnaprasad2023}, we add appropriate data noise \citep[following][]{yuan2012} and subsequently build time-distance maps to 
study the evolution of oscillations
along the loop. The oscillations are found only in three channels, the 171{\,}{\AA}, 193{\,}{\AA}, and 211{\,}{\AA}. The propagation properties are
exactly the same as those found in \citet{krishnaprasad2023} so we are not going to discuss them in detail here. Just highlighting one crucial point,
\citeauthor{krishnaprasad2023} found that the forward modelled propagation speeds in the 211\AA{} and 193\AA{} filter were very close to each other,
despite their difference in temperature. Here we explain quantitatively this phenomena: as can be seen in Table~\ref{tab3}, the predicted
propagation speeds $V$ are indeed very close to each other for these filters. We propose that the variation of phase speeds in different
observational filters \citep[e.g.][]{king2003} may be quantitatively explained through the proposed formula in Eq.~\ref{eq:filterspeed}: indeed, 
that equation shows that the observed phase speed $V$ is influenced by the specific filter's $\sigma_\mathrm{F}$ and $v_\mathrm{F}$. \par
\begin{figure}
	\centerline{\includegraphics[width=\columnwidth]{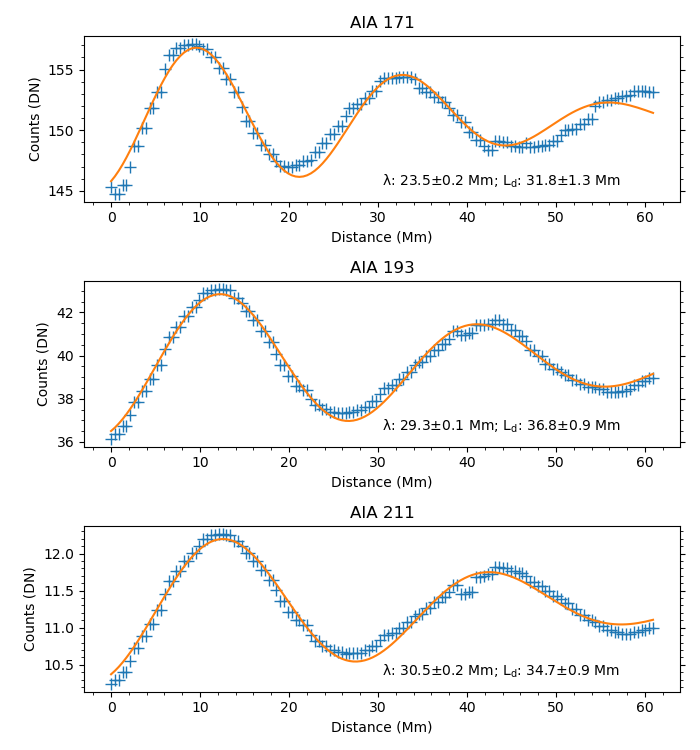}}
	\caption{The spatial intensity profiles at a particular instant along the loop obtained from the synthetic data corresponding to the AIA
171{,}{\AA}, 193{,}{\AA}, and 211{,}{\AA} filters. The solid lines represent a Gaussian damped sinusoid fit to the data following the function given
in Eq.{\,}\ref{eq:gdamp}. The obtained wavelength ($\lambda$) and damping length $L_d$ values are listed in the plot.}
	\label{fig:fomo_dlen}
\end{figure}
Let us now focus on the damping properties of the slow waves. Fig.{\,}\ref{fig:fomo_dlen} displays the spatial intensity profiles at a particular
instant along the loop for the three AIA channels. The solid lines in the figure correspond to the fitted profiles following Eq.{\,}\ref{eq:gdamp}. We 
notice that the fitted curve in AIA 171 has a significant deviation beyond a distance of 40{\,}Mm. Imposing tighter constraints on 
the fitting function did not improve the results much. However, the larger uncertainty obtained on the corresponding damping length should have 
incorporated this. The resulting associated damping lengths are 31.8$\pm$1.3{\,}Mm, 36.8$\pm$0.9{\,}Mm, and 34.7$\pm$0.9{\,}Mm, for the 171{\,}{\AA}, 
193{\,}{\AA}, and
211{\,}{\AA} channels, respectively. As may be noted these values are different from those obtained for the density and velocity parameters (see
Fig.{\,}\ref{fig:dlen}). This is due to the temperature response of the observing filter, which also has an influence on this multithermal 
apparent
damping, as described in Sec.~\ref{sec:filter}. \par

In order to quantitatively assess the effect of SDO/AIA filters, we fit the temperature response curves (version 9) for each coronal filter with a 
Gaussian function and estimate their standard width. 
\begin{figure*}
	\centerline{\includegraphics[width=\textwidth]{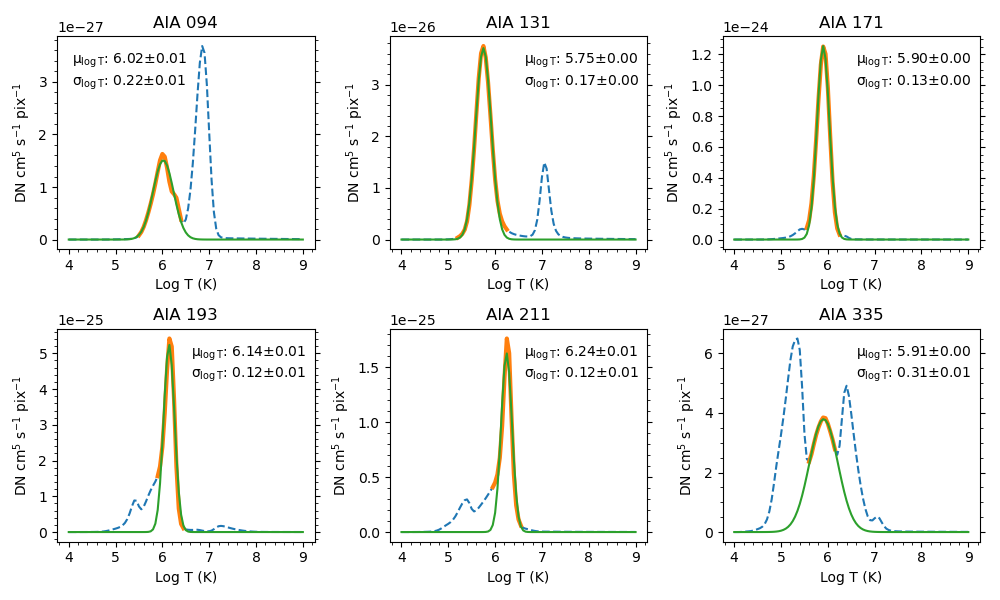}}
	\caption{SDO/AIA temperature response curves for the 6 coronal channels as listed. In each of the panels, the blue dashed line represents the full
response curve, the light orange solid line represents the segment fitted with a Gaussian function, and the green solid line represents the fitted 
function.
The obtained standard widths are listed in the plot. }
	\label{fig:filter_width}
\end{figure*}
The temperature response curves and the fitted profiles are plotted in Fig.{\,}\ref{fig:filter_width}. For the curves with multiple peaks, we choose
the peak that is closer to log{\,}$T=6.0$, which is the characteristic temperature in our simulations. In each of the panels, the dashed line shows
the full response curve, the light orange line denotes the fitted segment, and the green line shows the fitted function. The obtained widths
$\sigma_{\log{T}}$ are 0.22$\pm$0.01, 0.17$\pm$0.00, 0.13$\pm$0.00, 0.12$\pm$0.01, 0.12$\pm$0.01, and 0.31$\pm$0.01, for the 94{\,}{\AA},
131{\,}{\AA}, 171{\,}{\AA}, 193{\,}{\AA}, 211{\,}{\AA}, and 335{\,}{\AA} channels, respectively. These values along with the respective peak locations
are listed in Table~\ref{tab3}.  From these fitted filter curves in $\log{T}$-space, we have computed, for each filter, the corresponding peak
propagation speed $v_\mathrm{F}$ from the sound speed of the fitted peak temperature. Then, we have calculated $\sigma_\mathrm{F}$, as the average of
sound speeds belonging to the peak temperature plus and minus the filter peak width. Thus, we have assumed that the filter is symmetric in velocity
space. The results of these calculations are listed in table~\ref{tab3}. Then we have used Eqs.~\ref{eq:filterspeed} to compute $\Sigma$ and $V$, and
we have also listed the obtained values in table~\ref{tab3}.
\begin{table*}
\begin{center}
\caption{Properties of AIA filter response curves.}
\label{tab3}
\begin{tabular}{lcccccc}
\hline\hline
Channel name &AIA 94   & AIA 131 & AIA 171 & AIA 193 & AIA 211 & AIA 335 \\
Peak temperature $\mu_{\log{T}}$ ($\log{T}$) & 6.02$\pm$0.01 & 5.75$\pm$0.0 & 5.90$\pm$0.0 & 6.14$\pm$0.01 & 6.24$\pm$0.01 & 5.91$\pm$0.00 \\
$v_\mathrm{F}$ (km/s) & & & 135 & 179 & 200 & \\
Peak Width $\sigma_{\log{T}}$ ($\log{T}$) &0.22$\pm$0.01 & 0.17$\pm$0.0 & 0.13$\pm$0.0 & 0.12$\pm$0.01 & 0.12$\pm$0.01 & 0.31$\pm$0.01 \\
$\sigma_\mathrm{F}$ (km/s) & & & 20.3 & 24.7& 27.8 & \\
$\Sigma$ (km/s) & & & 16.1 & 18.0 & 19.1 & \\
$V$ (km/s) & & & 141.6 & 166.1 & 174.9 & \\
\hline
\end{tabular}
\end{center}
\end{table*}

For the density, we have calculated the predicted damping time with Eq.~\ref{eq:gaussdamp}. For predicting the damping times observed in the filters, 
we have used the values for $\Sigma$ and $V$ and inserted them in Eq.~\ref{eq:gaussobs}. The predicted 
damping values are listed in table~\ref{tab2}, along with the measured damping values from the simulations. The predicted damping times match
reasonable well with the modelled damping times (with a maximum deviation of 30\%). As before, we think that this deviation between the numerical
damping lengths and the predicted damping lengths is due to the finite number of strands of which the loop consists in the simulation. Thus, the
number of strands is insufficient to fill the entire Gaussian DEM. In essence, there are insufficient Monte Carlo realisations of the strands to
completely cover the expected Gaussian DEM distribution.
\\

\begin{table*}
\begin{center}
\caption{Gaussian damping lengths in Mm for various quantities.}
\label{tab2}
\begin{tabular}{cccccc}
\hline\hline
			   & Density & Velocity & AIA 171 & AIA 193 & AIA 211 \\
%Monolithic & 88.8$\pm$ 13.0& 91.1$\pm$14.0 & 88.7$\pm$10.7 & 87.9$\pm$10.5 & 87.6$\pm$10.2\\
Numerical model & 20.0$\pm$0.3& 21.2$\pm$0.3 & 31.8$\pm$1.3 & 36.8$\pm$0.9 & 34.7$\pm$0.9 \\
Predicted damping & 25.1 & & 35.7 & 43.8 & 45.8 \\
\hline
\end{tabular}
\end{center}
\end{table*}

\section{Conclusions and discussion}
In this paper, we have considered the apparent damping of slow waves (which we call ``multithermal apparent damping'' (MAD) or ``Voitenko-damping'') 
due to a
different propagation speed in coronal loop strands. We have considered a superposition of $\delta$-function impulses, Gaussian pulses or driven
waves. All of these models led to the multithermal apparent damping of slow waves due to observational phase mixing. We should stress
that the damping is indeed only apparent, and that no wave energy was \st{harmed} dissipated during the production of this paper. This multithermal 
apparent
damping of the slow waves is expected to be stronger than damping by thermal conduction for short periods (less than 200s), and comparable for longer
periods. We have found that the case of driven slow waves leads to a predicted Gaussian damping profile, with a predicted damping length $L_G$ of
\begin{equation*}
	L_G=\frac{\bar{v}^2}{\sigma_v\omega},
\end{equation*}
where $\bar{v}$ is the average sound speed in the loop, $\sigma_v$ is the spread in the sound speed, and $\omega$ is the frequency.
The resulting, predicted value of the damping length matches reasonably well with the one found in the simulations of \citet{krishnaprasad2023}. 
The predicted damping length scales linearly with the period of the wave. This is compatible with the observational synthesis made by 
\citet{cho2016}, who observed a unified picture of solar and stellar quasi-periodic pulsations with a damping time scaling linearly with the 
period. Moreover, this different scaling of the multithermal apparent damping time with period from thermal conduction may explain the difference 
in damping scalings in open-field or closed-field regions \citep[e.g.][and follow-up works]{krishnaprasad2014}, but also for different damping 
regimes at different heights \citep{gupta2014}. These different damping regimes could then be associated with different levels of multi-thermal 
structuring
of loops/plumes and the relative importance of thermal conduction damping and multithermal apparent damping.\par

In the second part of the paper, we have considered the effect of a finite filter width in imaging instruments such as SDO/AIA. We have found that
the finite filter has as effect that the waves have a different propagation speed $V$ and damping length $L_G$ in each filter. These are given by
\begin{equation*}
	V=\frac{\sigma_v^2v_\mathrm{F} + \sigma_\mathrm{F}^2
\bar{v}}{\sigma_\mathrm{F}^2+\sigma_v^2}, \quad L_G=\frac{V^2}{\Sigma\omega}, \quad
\frac{1}{\Sigma^2}=\frac{1}{\sigma_\mathrm{F}^2}+\frac{1}{\sigma_v^2}
\end{equation*}
where $v_\mathrm{F}$ is the central sound speed of the filter, and $\sigma_\mathrm{F}$ is the width of the filter. This explains two phenomena: (1)
the observed phase speed in different filters depends on the thermal properties of the loop, and (2) the damping in each filter is also different. We
have once again checked these formulas against the damping in forward models of the simulations of \citet{krishnaprasad2023}. We found that our
predictions match reasonably well with the simulated values (within 30\%). We suspect that the deviation is mostly caused by the small number of
strands in the simulation, which compares to our continuous DEM distribution that we considered in this paper.\par

We expect that these results may be used in the future to perform MHD seismology \citep{nakariakov2005} of coronal loops with slow waves. With the
above formulas, it is possible to fit the loop's DEM properties of central temperature (through the average sound speed $\bar{v}$) and spread in
temperature (through the value of the spread in sound speed $\sigma_v$). These DEM properties of the loops are only sensitive to the loop itself
in which the slow wave propagates. This is in contrast to the currently used method of DEM inversion
\citep[e.g.][]{hannah2012,cheung2015,krishnaprasad2018}, which is very sensitive to the careful background subtraction from the loop's emission. This
proposed method will at least allow to remove this sensitivity, and perhaps reveal more detailed thermal properties of loops. \\ Also the combination 
of spectral observations with imaging observations is an interesting avenue to consider, because the spectral observations are much less impacted by 
the multithermal apparent damping, and the combination of this with the imaging observations would allow to disentangle physical damping from the 
multithermal  apparent
damping. 

Topics for future research as a follow-up to this work would be (1) to consider the effect of a combination of multithermal apparent damping and 
thermal 
conduction in a multistranded loop system, (2) to model the effect of multithermal apparent damping on standing sound waves in e.g. flaring loops 
\citep{wang2011,cho2016}, and (3) investigation of the usage of different lines-of-sight from different spacecraft (e.g. Solar Orbiter and SDO) to 
probe the inner multithermal structure of loops using multithermal apparent damping properties. 

\begin{acknowledgements}
      TVD was supported by the European Research Council (ERC) under the European Union's Horizon 2020 research and innovation programme (grant 
agreement No 
724326), the C1 grant TRACEspace of Internal Funds KU Leuven, and a Senior Research Project (G088021N) of the FWO Vlaanderen. The research benefited 
greatly from discussions at ISSI. TVD would like to thank Dipankar Banerjee, Vaibhav Pant and Krishna Prasad for their hospitality when visiting 
ARIES, Nainital in spring 2023. TVD would like to thank Roberto Soler for referring to the \citet{voitenko2005} paper at the AGU Chapman conference
in Berlin 2023. VP is supported by SERB start-up research grant (File no. SRG/2022/001687). AWH acknowledges the financial support of the Science and 
Technology Facilities Council (STFC) through Consolidated Grants ST/S000402/1 and ST/W001195/1 to the University of St Andrews and support from the 
European Research Council (ERC) Synergy grant `The Whole Sun' (810218). The computational resources and services used in this work were provided by 
the VSC (Flemish Supercomputer Center), funded by the Research Foundation Flanders (FWO) and the Flemish Government – department EWI.
\end{acknowledgements}

% WARNING
%-------------------------------------------------------------------
% Please note that we have included the references to the file aa.dem in
% order to compile it, but we ask you to:
%
% - use BibTeX with the regular commands:
\bibliographystyle{aa} % style aa.bst
\bibliography{refs} % your references Yourfile.bib
%
% - join the .bib files when you upload your source files
%-------------------------------------------------------------------
\end{document}